\documentstyle[12pt]{article}
\newcommand{\beq}{\begin{equation}}
\newcommand{\eeq}{\end{equation}}

\begin{document}
\def\lag{\langle}
\def\rag{\rangle}
\begin{titlepage}
\vspace{2.5cm}
\begin{center}
\Large\bf{\hfill\break
          \hfill\break
          \hfill\break
Multicanonical Study of the 3D Ising Spin Glass$\,^{1}$}\\
\vspace{2.5cm}
\large{Bernd A. Berg,$^{2,3}$
Tarik Celik$\,^{2,4}$
and Ulrich Hansmann$\,^{2,3}$
}
\end{center}
\vspace{6cm}
\begin{center}
{\bf Abstract:}
\end{center}

We simulated the Edwards-Anderson Ising spin glass model in three
dimensions via the recently proposed multicanonical ensemble. Physical
quantities such as energy density, specific heat and  entropy are evaluated
at all temperatures. We studied their finite size scaling, as well as the zero
temperature limit to explore the ground state properties.
\vfill

\footnotetext[1]{{This research project was partially funded by the
Department of Energy under contracts DE-FG05-87ER40319,
DE-FC05-85ER2500, by FSU's COFRS program, by the Deutsche
Forschungsgemeinschaft \hbox{under} contract H180411-1,
and by the NATO Science Program. Submitted to
                      {\it Zeitschrift f\"ur Physik B}.}}
\footnotetext[2]{{Supercomputer Computations Research Institute,
                      Tallahassee, FL~32306, USA.}}
\footnotetext[3]{{Department of Physics, The Florida State University,
                      Tallahassee, FL~32306, USA. }}
\footnotetext[4]{{On leave of absence from Department of Physics, Hacettepe
                  University, Ankara, Turkey.}}
\end{titlepage}

One of the great challenges in numerical simulations is the study of
spin glasses. These systems display a rather peculiar structure due to
the effects of disorder and frustration, for reviews see \cite{Bi1,book}.
Recently two of the authors explored the multicanonical ensemble
\cite{our1} for simulations of spin glasses. Results for the 2D
Edwards-Anderson Ising spin glass  were encouraging \cite{our3}. Similar
concepts were also tested in \cite{Enzo}. In this letter we present a
preliminary multicanonical simulation of the physically more interesting
3D Edwards-Anderson Ising spin glass.

The Hamiltonian of the model is given by
$$ H\ =\ - \sum_{<ij>} J_{ij} s_i s_j , \eqno(1) $$
where the sum goes over nearest neighbours and the exchange interactions
$J_{ij}=\pm 1$ between the spins $s_i=\pm 1$ are randomly distributed over
the lattice with the constraint $\sum J_{ij} = 0$ for each realization.
Despite its simplicity the model is supposed to be sufficiently realistic.
Recent simulations \cite{Cara} of the 3D model in a magnetic field favour
the mean field picture rather than the alternative droplet model
\cite{Huse,Moore}. However, it can be argued that equilibrium at
sufficiently low temperatures has not been reached \cite{Fish}. For previous
simulations of Ising spin glasses in 3D see \cite{Bhatt,Ogie,Bray,McMi,Kirk}.

The multicanonical ensemble \cite{our1} can be defined by weight factors
$$
P_M (E)\ =\ \exp \left[ -\beta (E) E + \alpha (E) \right] .\eqno(2)
$$
$\alpha (E)$ and $\beta (E)$ are to be determined such that for the
chosen energy range $E_{\min}\le E\le E_{\max}$ the resulting
multicanonical probability density is approximately flat:
$$
P_{mu} (E)\ =\ c_{mu}\ n(E) P_M (E)\ \approx\ \hbox{const} \eqno(3)
$$
where $n(E)$ is the spectral density.
In the present study we take $E_{\max} =0$ ($\beta (E) \equiv 0$ for
$E\ge E_{\max}$) and $E_{\min}=E^0$ the ground state energy of the
considered spin glass realization.
The purpose of the function $\alpha (E)$ is to give $\beta (E)^{-1}$
the interpretation of an effective temperature. This leads to the
recursion relation
$$
\alpha (E-4)\ =\ \alpha (E) +
\left[ \beta (E-4) - \beta (E) \right] E,\
\alpha(E_{\max}) = 0  . \eqno(4)
$$
The multicanonical function $\beta (E)$ is obtained  via recursive
multicanonical calculations. One performs simulations with $\beta^n (E)$,
$n=0,1,2,\,$\ldots, which yield probability densities $P^n (E)$ with
medians $E^n_{\tenrm median}$.  We start off with $n=0$
and $\beta^0 (E) \equiv 0$. The recursion from $n$ to $n+1$ reads
$$ \beta^{n+1} (E)\ =
\left\{
   \begin{array}{ll}
   	 \beta^n (E)\   \hbox{for} \ E \ge E^n_{\tenrm median};& \\
\\
   	 \beta^n (E)\ + 0.25\times\ \ln \left[ P^n(E+4)/P^n(E) \right]\\
\qquad\qquad	 \hbox{for\ \ }   E^n_{\tenrm median} > E \ge E^n_{\min}; &\\
\\
   	 \beta^{n+1} (E^n_{\min})\   \hbox{for\ \ }  E < E^n_{\min}\, .&{}
\end{array}
\right.
\eqno(5) $$
The recursion is stopped for $m$ with
$E^{m-1}_{\min}=E^0$ being groundstate.

By weighting with $\exp [-\hat\beta E + \beta (E) E - \alpha (E)]$
canonical expectation values
$ {\cal O} (\hat\beta)\ =\ Z({\hat\beta} )^{-1}
\sum_E {\cal O} (E) n(E) \exp (-\hat\beta E) $, where
$ Z(\hat\beta )\ =\ \sum_E n(E) \exp (-\hat\beta E) $
is the partition function, can be reconstructed for all $\hat\beta$
(the canonical inverse temperature).
The normalization constant $c_{mu}$ in equation (3)
follows from $Z(0)=\sum_E n(E) = 2^N$, where $N$ is the total number of
spin variables. This gives the spectral density and allows to calculate
the free energy as well as the entropy.

We simulated the Edwards-Anderson Ising spin glass model
on 3D cubic lattices with linear size $L=$ 4, 6, 8 and 12.
We performed simulations of $2\times 10^6$ iterations on $4^3$ lattice
with $32$ different realizations of random variables $J_{ij}$.
We carried out up to $4\times 10^6$ iterations on  $4^3$, $6^3$ and
$12^3$ lattices with 16, 8 and 4 realizations, respectively. Thermal
averages were evaluated after the first additional $2\times 10^5$
iterations being discarded, although with a disordered starting
configuration the multicanonical ensemble is immediately in equilibrium.

Our results for physical quantities are summarized in Table 1. The final
mean values and their error bars are obtained by combining the
results from the different realizations. Different realizations are
statistically independent and enter with equal weights. The final
error bars are  enlarged by a Student multiplicative factor, such
that the probability content of two standard deviations is
Gaussian (95.5\%).

\begin{table}[h]
\centering
\begin{tabular}{||c|c|c|c|c|c|c|c||}                    \hline
$          $  & $L = 4$   & $L = 6$ & $L = 8$  & $L = 12$  \\ \hline
$\beta_{\max}$&1.04 $\pm$ 0.03&1.52 $\pm$ 0.12&2.04 $\pm$ 0.18&
              $2.30 \pm 0.38 $
\\ \hline
$\tau^e_{L}$ &685 $\pm$ 95 &30166 $\pm$ 16383&117038 $\pm$ 40630&
           $1580230 \pm 882686 $
\\ \hline
$e^0$ &$-$1.7403 $\pm$ 0.0114&$-$1.7741 $\pm$ 0.0074&$-$1.7822 $\pm$ 0.0081&
        $-$1.7843 $\pm$ 0.0030
\\ \hline
$s^0$   &0.0724 $\pm$ 0.0047&0.0489 $\pm$ 0.0049&0.0459 $\pm$ 0.0030 &
        $0.0491 \pm 0.0023 $
\\ \hline
$\chi^0_q$ &0.65 $\pm$ 0.05 &0.71 $\pm$ 0.05 &0.56 $\pm$ 0.10 &
        0.68 $\pm$ 0.10
\\ \hline
\end{tabular}
\caption{\em}
\end{table}

The energy density $e(\hat\beta )$, the specific heat $c(\hat\beta )$,
and the entropy per spin $s(\hat\beta )$ follow in a
straightforward manner by constructing the canonical ensemble.
Fig.~1 depicts the energy density and the entropy per spin
versus inverse temperature for $L=8$. The indicated error bars are
with respect to the eight different realizations. For the
temperature range $0.8 \le T \le 1.6$ we collect in Fig.~2 our
$L=4, 6, 8$ and 12 results for the specific heat.
This is of interest as a spin glass phase transition is
claimed to take place at $T_c \approx 1.2$ \cite{Bhatt,Ogie}.
However, this picture was recently questioned by Bhanot and Lacki
\cite{Bhan} on the basis of exact partition function calculations
on fairly small systems (up to $5\times 5\times 9$). The critical
exponent estimate $\nu = 1.3$ \cite{Bhatt,Ogie} implies
$\alpha =-1.9$ via the hyperscaling relation $D\nu = 2-\alpha$.
The finite size scaling (FSS) of the specific heat $T=T_c$ is
{}~$ c \sim  L^{\alpha / \nu} $. Clearly, Fig.~2 is inconsistent with
this $L$-dependence and exhibits instead
almost negligible finite size effects as would be typical for
non-critical behaviour.

\begin{figure}
\vspace{6in}
\caption{Energy density and entropy per spin versus $\beta$ (from the $L=8$
lattices).}
\end{figure}

\begin{figure}
\vspace{3in}
\caption{Specific heat versus temperature for $L=4 (\times ), 6 ({\rm o}),
8 (\Box) $ and $12 (*)$ lattices.}
\end{figure}

\begin{figure}
\vspace{6in}
\caption{FSS estimates of the infinite volume groundstate energy density
and entropy per spin.}
\end{figure}

Let us now concentrate on the zero temperature $(\hat\beta\to\infty )$
limit. The groundstate energy density is $e^0 = E^0/N$,
and we obtain its entropy per spin as $s^0 = S^0/N = \ln [n(E^0)]/N$.
FSS fits of the form $f^0_L = f^0_{\infty} + c/N$ are used
to estimate the infinite volume groundstate energy density and entropy per
spin.
Fig.~3 displays these fits. Our groundstate energy density result
$e^0 = -1.7863 \pm 0.0028$ is consistent with the previous estimates
$e^0 = -1.76 \pm 0.02$ \cite{MorBi} and $e^0 = -1.75 $ \cite{Kirk}.
Our value for the groundstate entropy per spin $s^0 = 0.046 \pm 0.002$ is
consistent with $s^0 = 0.04 \pm 0.01$ given by Morgenstern and Binder
\cite{MorBi} and lower than Kirkpatrick's value $s^0 = 0.062$ which was
estimated from a single  sample of $20^3$ spins \cite{Kirk}. Our groundstate
entropy value translates, even for moderately sized systems, into
large numbers of distinct groundstates. For instance $s^0= 0.046$ implies
approximately $3.3 \times 10^{34}$ groundstates for a $12^3$ lattice.
We also tried the entropy fit $N s^0_L = \ln (N) + c$, which corresponds
to a power law ansatz for the groundstate degeneracy. Unacceptable
$\chi^2$ values rule out this fit even if the lattice range is restricted
to $L=4-8$.

To visualize the low temperature behaviour, we show in Fig.~4 the spin
glass order parameter distribution of one of our $L=8$ realizations. Its
calculation follows the lines of \cite{our3}. One clearly sees five
configuration space valleys which are
separated by high tunneling barriers. The multicanonical simulation
overcomes these energy barriers by connecting back to the disordered
high temperature states. The realization from which Fig.~4 is depicted
showed nine tunneling events. Here tunneling means that starting from the
high temperature disordered phase the algorithm finds a true ground state
and then travels all the way back to the disordered phase. The average
number of sweeps needed for this process is our ergodicity time $\tau^e_L$.
Table 1 shows how the ergodicity times increase with the lattice size for
our simulations. The data  corresponds in CPU time to a slowing down
$\sim V^{3.4 (2)}$ which is similar to the one involved in our previous
2D simulations.

Further, we have included in Table 1 our zero temperature (groundstates)
results $\chi^0_q$ for the spin glass susceptibility density
$\chi_q = <q^2> / V$ (note that $<q>=0$ due to the symmetry of the
Hamiltonian). The numbers are modestly indicative for a non-trivial
$P_0 (q)$ distribution (lack of self-averaging). We also looked at
the temperature dependence of the spin glass susceptibility and the
corresponding Binder parameter. In the neighbourhood of the
suspected $T_c$ our results are consistent with [10,11], but our
present accuracy does not allow a detailed finite size scaling
investigation. Considerable future improvements are feasible.

\pagebreak[4]

\begin{figure}
\vspace{4in}
\caption{Spin glass order parameter distribution ($L=8$ lattice).}
\end{figure}

Let us state the conclusions. Our results on specific heat,
shown in Fig.~2, raise some additional doubts about the standard picture
of the spin glass phase transition. But on a qualitative level we find
a bifurcation of the spin glass order parameter, as depicted in Fig.~4.
A future, more quantitative, investigation of this quantity is desirable.
Fig.~4 and our preliminary results for the spin glass susceptibility
favour the mean field picture of spin glasses. To reach seminal
results one has to investigate larger lattices and, first of all,
to clarify the nature of the transition.
\hfill\break

We would like to thank Thomas Neuhaus and Mark Novotny for
discussions. B.A.B. acknowledges discussions with Kurt Binder.
T.C. was supported by TUBITAK of Turkey and U.H by the Deutsche
Forschungsgemeinschaft. Both would like to thank Joe Lanutti and SCRI for the
warm hospitality extended to them. Our simulations were performed on
the SCRI cluster of RISC workstations.
\vfill \hfill \eject

\eject

\end{document}